# Transient Phase Sensing in a Three-Photon Rydberg Ladder Scheme

Stephanie M. Bohaichuk[1], Vijin Venu[1], Florian Christaller[1], James P. Shaffer[1]

[1]Quantum Valley Ideas Laboratories, 485 Wes Graham Way, Waterloo, ON N2L 0A7, Canada

*Although Rydberg atoms have shown promise for use in novel types of radio frequency receivers, they have generally not been considered phase sensitive without the use of closed-loop interferometry or auxiliary radio frequency fields. Here, we show that the high coherency of a narrow-linewidth three-photon ladder excitation scheme unique to Cesium atoms enables all-optical sensing of transient changes in RF phase within a room temperature vapor cell. The transient response on the probe laser's transmission originates from phase-to-amplitude conversion via a disturbance of the coherency of the system in response to the phase shift of the radio frequency field. We show that the amplitude and frequency of the oscillatory response provides information on the magnitude and direction of any radio frequency field detuning. We demonstrate that the detuning sensitivity can be used to identify Doppler shifts in radar applications, by applying phase shifts embedded in radio frequency pulses. The phase modulation within the radar pulse acts as a form of compression that facilitates the simultaneous detection of both target position and velocity.*

Rydberg atoms are at the heart of a variety of developing quantum technologies, including as detectors of radio frequency (RF) electric fields (E-fields) for use in RF test and measurement, radar, and telecommunications [1]. When alkali atoms in a vapor cell, such as cesium or rubidium, are optically excited to Rydberg states generating electromagnetic transparency (EIT) or absorption (EIA), they become highly sensitive to the presence of RF electric fields that are (near) resonant with atomic transitions [2–4]. Changes in the RF E-field are read out optically by monitoring changes in a probe laser's transmission through the vapor. Information about the amplitude, frequency, and polarization of the RF E-field can be detected with high sensitivity [5–9], but phase detection, especially all-optical, has proven challenging.

Atomic RF sensors have generally not been considered phase sensitive, because the steady state response of the atomic system does not depend on the phases of the driving fields. Experimental phase detection has relied on use of an additional RF E-field, either as a RF local oscillator for heterodyning or as part of closed-loop interferometry [10–13]. Closed loop systems have been demonstrated using multiple RF E-fields [13], while optical closed loops with a single RF E-field have been discussed in theory but are challenging to implement in practice [14–16].

In this work, we show that a three-photon ladder excitation scheme senses changes in the phase of an incident RF E-field as transient oscillations in the probe laser's transmission, self-referenced to the steady state, eliminating the need for any atomic closed loop or RF mixing. The co-linear three-photon excitation scheme in cesium provides narrow linewidth and high coherence even at room temperature, because of the wavevector matching. Doppler shifts in the RF frequency can be determined using the oscillatory response to phase modulation, showing that Rydberg sensors can be utilized effectively as pulsed-Doppler receivers.

Our co-linear three-photon cesium excitation scheme and optical setup are shown in Figure 1a. The $42P_{3/2}$ Rydberg state is reached by an 895 nm probe laser driving the $6S_{1/2}(F=4) \rightarrow 6P_{1/2}(F=3)$ transition, a 636 nm intermediate laser driving the $6P_{1/2}(F=3) \rightarrow 9S_{1/2}(F=4)$ transition, and a 2262 nm coupling laser driving the $9S_{1/2}(F=4) \rightarrow 42P_{3/2}$ transition [6]. The 636 nm laser counter-propagates with the 895 nm and 2262 nm lasers, reducing the wavevector mismatch and Doppler broadening, producing spectral full-width-half-maximum linewidths of $\sim 2\pi \times 220$ kHz under the conditions used here. The beam diameters of the 895 nm, 636 nm, and 2262 nm lasers are 4.8 mm, 5.2 mm, and 6.4 mm, respectively, with linewidths below $2\pi \times 0.6$ kHz, $2\pi \times 5$ kHz, and $2\pi \times 20$ kHz, respectively. All three lasers are Pound-Drever-Hall locked to ultra-low expansion Fabry-Perot cavities. We use a cylindrical glass-blown cesium vapor cell at room temperature with 2.54 cm length and diameter for the Rydberg atom sensor. Compensation coils around the vapor cell are used to cancel the Earth's magnetic field, in order to limit Zeeman shifts of the atomic states, making interpretation of the experiment more straightforward.

We use a 10.7 GHz RF wave emitted by a horn antenna ~30 cm from the vapor cell, which propagates perpendicular to the optical axis and is vertically polarized. A phase shift is generated on the RF wave by an RF generator. The RF

E-field is (near) resonant with the $42P_{3/2} \leftrightarrow 41D_{5/2}$ transition, for which we calculated a dipole moment of $1082\ ea_0$ when optical pumping in the full hyperfine basis is considered for the polarizations and states used for the experiments [9].

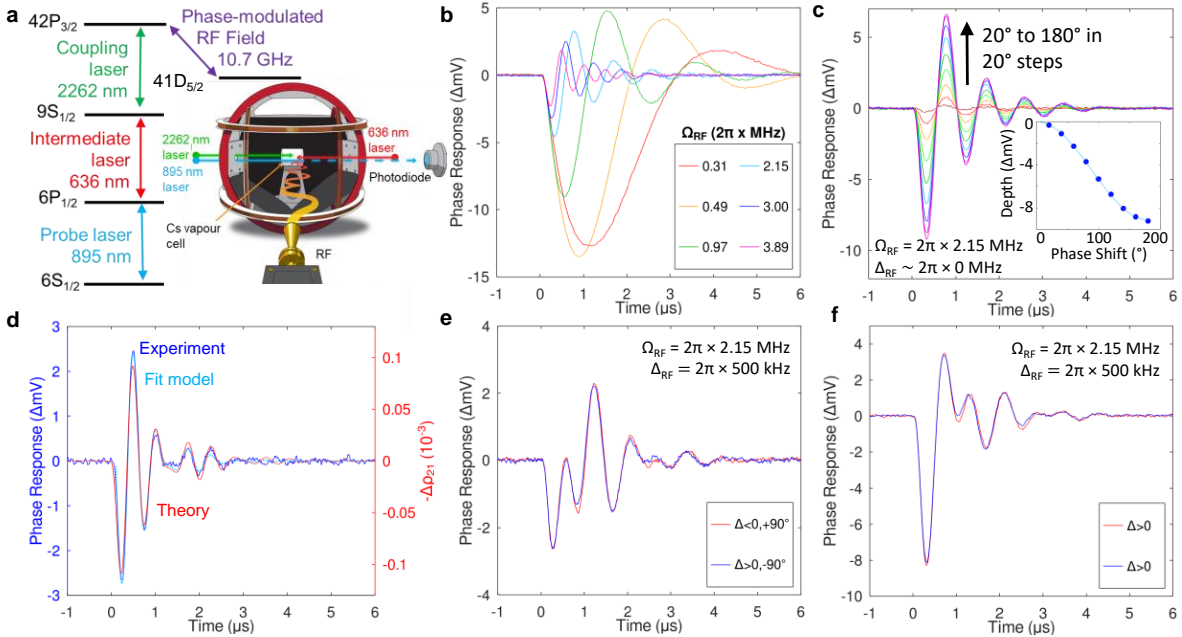

FIG. 1 (a) 3-photon ladder excitation scheme and experimental setup used. An abrupt change in phase of the RF E-field emitted by the horn is detected via the probe laser's transmission through a 2.5 cm cesium vapor cell. (b) A phase jump in an RF E-field produces an oscillatory response with a decaying exponential envelope. The oscillation frequency is directly related to the RF Rabi frequency ($\Omega_{RF}$), i.e. the E-field amplitude. The RF is on resonance, with an abrupt +90° phase step. (c) On resonance, the amplitude of the phase response scales nonlinearly with the phase shift magnitude ($\Delta\theta$) but the decay and frequency of oscillations remains constant. The inset shows how the depth of the first minimum scales with $\Delta\theta$, with a fit to $C_0(1-\cos(\Delta\theta))$ shown in blue. (d) Experimental atomic response to a +90° RF phase shift (dark blue) compared to one simulated with a density matrix model (red). The light blue dashed line is a fit to the experimental data using Eqn. (1). $\Delta_{RF} = 2\pi \times 0.60$ MHz. (e) Reversing the sign of the RF detuning is equivalent to reversing the sign of the phase shift. (f) If the phase shift is 180° then the response is invariant to the sign of the detuning. $\Omega_{895} = 2\pi \times 0.2$ MHz, $\Omega_{636} = 2\pi \times 3.4$ MHz, $\Omega_{2262} = 2\pi \times 0.2$ MHz.

The atomic response to a +90° phase jump on a resonant RF E-field is shown in Figure 1b, consisting of decaying oscillations around steady-state. Sudden changes in the RF phase perturb the coherence of the system which results in phase information being converted by the atoms to amplitude oscillations in the probe laser's transmission. When the RF phase changes, the field vector on the Bloch sphere of the RF transition rotates in the x-y plane. Since the dressed states of the original system lie parallel and anti-parallel to the field vector, the state vector is no longer in steady-state after the phase change, leading to relaxation of the system to the new steady-state, i.e. alignment with the new field vector.

The three-photon scheme used in this work provides high sensitivity to transient phase shifts in the RF, due to the inherent narrow linewidth and associated high coherence time. We measure an EIA linewidth of $\Gamma = 2\pi \times (222 \pm 6)$ kHz when the 2262 nm coupling laser is scanned, implying a coherence time on the order of $\tau \sim \Gamma^{-1} = 0.7$ μs that allows several Rabi oscillation cycles to be visible before being damped out. The high coherence highlights that the coupling laser fields interacting with the atom serve to dress it, effectively changing its properties so as to respond in a desirable manner to the incident radio frequency field. The strongest amplitude of phase response generally occurs at an RF Rabi frequency comparable to $2\Gamma$. Transient oscillations occur with a higher frequency at higher RF Rabi frequency ($\Omega_{RF}$), albeit at a reduced amplitude. Increasing the magnitude of the phase shift ($\Delta\theta$) increases the overall amplitude of the transient phase response, shown in Figure 1c. On resonance, the amplitude scales non-linearly in proportion to $C_0(1-\cos(\Delta\theta))$, where $C_0 = -4.56$ mV is an amplitude fit to data in Figure 1c (blue line) that will depend on the laser Rabi frequencies.

Transient dynamics in atomic systems generally consist of exponentially decaying sinusoidal components [17–20]. In an isolated two-level system these are given by Torrey's solutions, which have a damped oscillation at the generalized Rabi frequency driving the transition. In our system, we find that the dynamics are largely determined by damped Rabi oscillations originating on the two-level RF transition that are transferred to the probe laser. Unlike the two-level system, the RF transition is not isolated from the rest of the ladder; in particular, the lower Rydberg state is optically pumped by the coupling laser and both Rydberg states decay to other levels. We find that this modifies the two-level dynamics to produce two oscillation frequencies, which are given by the eigenenergies of the two Autler-Townes dressed states [21,22]:

$$\omega_{1,2} \sim \frac{1}{2}\left(\Delta_{RF} \pm \Omega_{RF}^g\right) \qquad (1)$$

where $\Omega_{RF}^g = \sqrt{\Omega_{RF}^2 + \Delta_{RF}^2}$ is the generalized Rabi frequency and $\Delta_{RF}$ is the detuning of the RF E-field from atomic resonance. Empirically we find the change in probe transmission ($\Delta T$) is approximately given by a sum of these two oscillations:

$$\Delta T \approx \exp(-t/\tau)\left[A\cos(\omega_1 t + \phi_1) + B\cos(\omega_2 t + \phi_2)\right] \qquad (2)$$

$\tau$ is the coherence time determined by the system's decay and dephasing rates, which typically include transit time broadening, laser linewidths, and Rydberg-Rydberg atom collisions which contribute to dephasing.

As the RF detuning nears zero we expect that the two oscillation frequencies approach degeneracy at $\omega_{1,2} \sim \Omega_{RF}/2$, with $A \sim B$ and $|\phi_1 - \phi_2| \sim 0°$. Density matrix simulations near resonance confirms the degeneracy in the absence of Doppler broadening. In the presence of a distribution of RF detunings, which arises from Doppler broadening, or a distribution of RF amplitudes, which can originate from scattering of the RF field within the vapor cell, deviations from degeneracy of the oscillation frequencies can be observed both the calculations and experiment.

A typical least squares fit of Eqn. (2) to experimental data is shown as a light blue dashed line in Figure 1d. The measured continuous wave (CW) RF E-field amplitude from Autler-Townes splitting is $\Omega_{RF} = 2\pi \times 3.89$ MHz and the known RF detuning is $\Delta_{RF} = 2\pi \times 600$ kHz, resulting in expected oscillation frequencies of $2\pi \times 2.27$ MHz and $2\pi \times 1.66$ MHz. The fit obtains parameters $A = 2.07 \pm 0.03$ mV, $B = 3.39 \pm 0.06$ mV, $\tau = 0.631 \pm 0.008$ μs, $\phi_1 = 4.93 \pm 0.03$ rad, $\phi_1 = 1.23 \pm 0.01$ rad, $\omega_1 = 2\pi \times (2.340 \pm 0.005)$ MHz and $\omega_2 = 2\pi \times (1.65 \pm 0.003)$ MHz. At finite RF detunings there will be different absorption strengths and populations for each of the Autler-Townes dressed states due to optical pumping, and as a result the oscillation amplitudes $A$ and $B$ will not be equal.

A comparison with numerical results of a density matrix model calculation is also shown in Figure 1d in red. The model solves the time-dependent master equation of the five-level ladder system, which includes a phase step in the RF Rabi frequency, and thermally averages over velocity classes. We include radiative and blackbody decay rates $\Gamma_{21} = 2\pi \times 4.5$ MHz and $\Gamma_{32} = 2\pi \times 200$ kHz, as well as a transit time broadening of $2\pi \times 50$ kHz. Dephasing rates for each transition are added based on the sum of the laser spectral linewidths interacting with each level. An additional dephasing of $2\pi \times 175$ kHz is added to the RF transition, to account for spectral broadening seen in the Autler-Townes peaks, primarily attributed to RF inhomogeneity within the vapor cell. The numerical model shows excellent agreement with experiment, and fully captures the transient behaviour of the system.

Figures 1e-f highlight several symmetries of the phase response. Flipping the sign of a |90°| phase shift produces the same response as having done the measurement at an oppositely signed RF detuning, shown in Figure 1e. A phase shift in a two-level system can be viewed as a rotation in the x-y plane of the driving field vector on the Bloch sphere. A rotation due to a positive phase shift of a driving field with a positive detuning, i.e. a positive tilt out of plane, will produce a new driving vector that is exactly opposite that from a rotation in the opposite direction with a driving field that has a negative tilt out of plane. Given the symmetry of the shift and the same final axis of procession, both produce the same dynamics. At a phase shift of |180°|, the phase response is the same for both signs of RF detuning.

Figures 2a-b shows how the transient atomic responses to a +90° and -90° phase jump in the RF E-field change as the RF E-field is detuned increasingly off resonance. We note that "resonance" here refers to the five-level atomic system as a whole. That is, if one of the lasers is detuned then the apparent resonant point of the RF shifts to compensate. The three-photon system can be sensitive to kHz level detunings that are not apparent in coarser spectroscopy like saturated absorption and the two-photon EIT used in conventional Rydberg sensors [2]. Care must be taken to assure the overall system is resonant, or that fixed detunings of the lasers have been achieved. The same magnitude of detuning produces a more dramatic change in the transient phase response at weaker RF Rabi frequencies, because it takes a much smaller $\Delta_{RF}$ to become comparable to $\Omega_{RF}$ in Eqn. (1).

The transient phase response of the atoms contains information about an incident RF E-field amplitude, its frequency, and sub-microsecond arrival time information. One potential application is in radar systems, where a shift in the frequency of the RF wave reflected off a moving target identifies its velocity via the Doppler shift. The table in Fig. 2c converts several example velocities to equivalent RF detunings. An ideal radar receiver detects both the position and velocity of a target simultaneously by using Doppler shifts. We find that phase modulation added on top of RF radar pulses can be used to detect small RF frequency deviations. Changing the sign of the phase shift or the direction of detuning changes the relative phases of the two oscillatory components comprising the transient response. This produces an asymmetry between +90° and -90° responses that can be quickly quantified via the depth of the first oscillation (generally, the global maximum or minimum). This is shown in Figure 2d and can be used to read out $\Delta_{RF}$, especially at small $\Delta_{RF}$ where fitting to Eqn. (2) is less reliable because $\omega_1 \sim \omega_2$. A 90° magnitude of phase shift is found to produce the highest sensitivity to detuning. The asymmetry is more robust to fluctuations in laser power than using the absolute depth of a single phase shift's response. At small $\Delta_{RF}$, the asymmetry varies approximately linearly with $\Delta_{RF}$, with a higher slope at weaker RF E-field amplitudes. RF detunings of ~10 kHz can clearly be distinguished, as seen in Figure 2b. If the linewidth of the Autler-Townes peaks can be reduced, increasing the coherence time $\tau$ and the amplitudes $A$ and $B$, then we expect smaller detunings to be resolved. For the states used in this work, a theoretical minimum linewidth of the 3-photon EIA is expected to be around $2\pi \times 50$ kHz if laser linewidths and transit time broadening are minimal, and smaller if adiabatic elimination is used [23]. Achieving this would result in a ~6× improvement in $\tau$ and therefore at least a ~2× increase in oscillation depth due to the reduced exponential damping alone, resolving detunings on the ~kHz level or below.

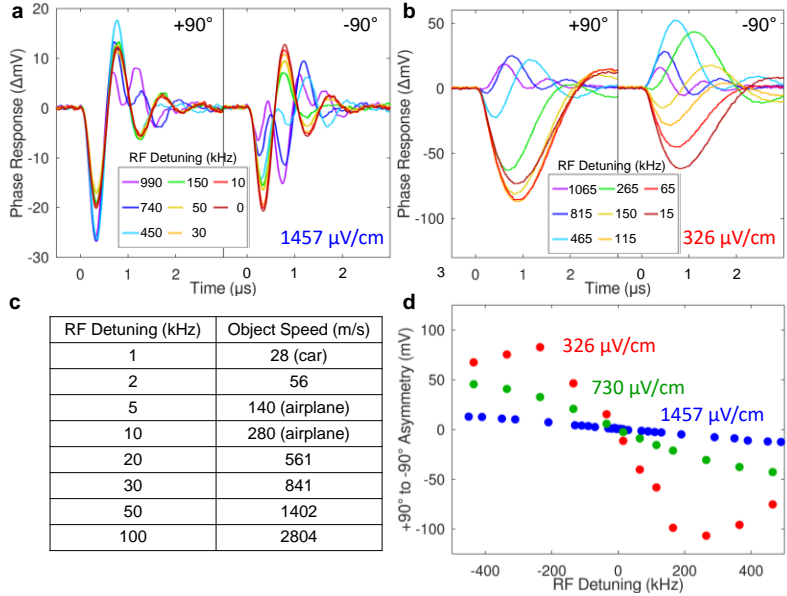

FIG. 2 (a)-(b) Averaged atomic response to an RF phase shift of +90° followed 10 μs later by -90°. As the RF is detuned off resonance, an asymmetry develops between the two directions of phase shift, changing the phase and amplitude of the oscillatory components making up the response. $\Omega_{895} = 2\pi \times 0.1$ MHz, $\Omega_{636} = 2\pi \times 2.4$ MHz, $\Omega_{2262} = 2\pi \times 0.2$ MHz. (c) Conversion between RF detuning amounts (Doppler shifts) and target speeds. (d) Change in the asymmetry of the atomic phase response to opposite phase shift directions at different detunings. We define asymmetry as the change in the depth of the first minimum of the oscillation between a +90° and a -90° phase shift response.

We also observe damped oscillatory behavior on the leading edge of the atomic response to a square RF pulse without any phase modulation, shown in Figure 3a. The introduction of an RF field to the system causes a similar perturbation resulting in damped Rabi oscillations determined by $\Delta_{RF}$ and $\Omega_{RF}$, with oscillation frequencies given by Eqn. (1). The high coherency of the three-photon system allows these oscillations to be clearly visible for several cycles before decaying, unlike in RF pulses measured with the standard cesium two-photon ladder scheme which is dominated by Doppler broadening [24].

Using the changes in the frequency components of the oscillations, or the degree of correlation with an associated matched filter template, the RF detuning can be determined from a standard square RF pulse without the need for additional modulation. This enables monitoring of velocity during rapid short pulse trains in radar, provided each pulse is longer than at least one Rabi oscillation. Such trains can be used to provide high positional precision and range resolution, but in scenarios involving weak signals or low energy emitters long pulses are preferred for improved signal-to-noise. Adding a phase modulation effectively compresses target velocity information into a long pulses, an example of which is shown in Figure 3b-c. Alternating +90° and -90° phase shifts are added at a time interval of 8 μs to the square amplitude RF pulse. The frequencies of the phase oscillations can further provide information about both RF amplitude and frequency. Asymmetry between the two phase shifts can identify RF detunings below 10 kHz. Timing information of single pulses can still be obtained using the peak of the output of matched filtering, shown in Figure 3d. The addition of the phase modulation does not affect the overall shape of the matched filter output compared to a standard square RF pulse.

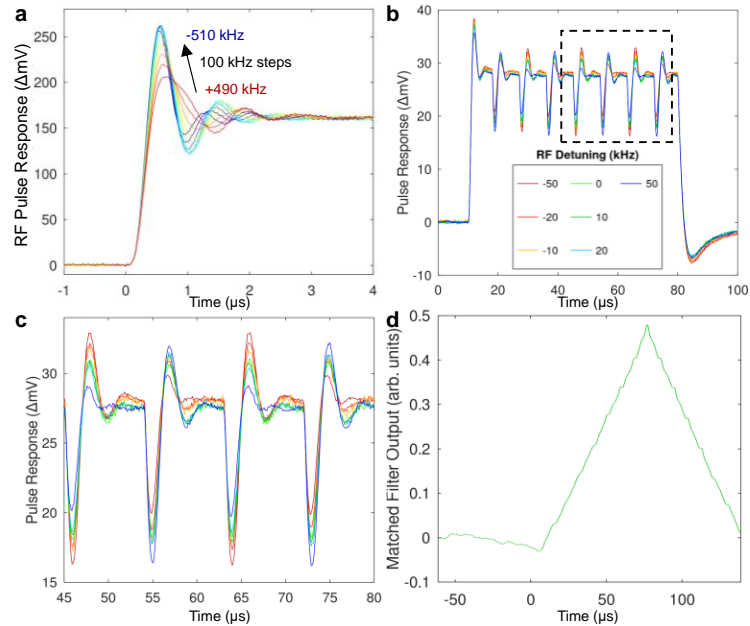

FIG. 3 (a) Leading edge of the atomic response to an RF pulse at different RF detunings, which also contains decaying Rabi oscillations at the same frequencies as the transient phase response of the atomic system. RF detunings span from +490 kHz (dark red) through the rainbow to -510 kHz (dark blue) in steps of 100 kHz. $\Omega_{895} = 2\pi \times 0.1$ MHz, $\Omega_{636} = 2\pi \times 2.4$ MHz, $\Omega_{2262} = 2\pi \times 0.2$ MHz, and $\Omega_{RF} = 2\pi \times 2.0$ MHz. (b) Example averaged radar-like RF pulse with phase modulation providing RF frequency information via phase response asymmetry, shown in more detail in (c). (d) Processing a single radar-like RF pulse with a resonant matched filter can identify arrival time of a radar echo using the peak of the response with no distortion from the phase modulation. $\Omega_{895} = 2\pi \times 0.2$ MHz, $\Omega_{636} = 2\pi \times 3.4$ MHz, $\Omega_{2262} = 2\pi \times 0.2$ MHz, and $\Omega_{RF} = 2\pi \times 0.5$ MHz.

In conclusion, we experimentally demonstrate that the coherency of a co-linear three-photon excitation scheme in a cesium vapor cell enables detection of transient Rabi oscillations in response to changes in amplitude or phase of an incoming RF E-field. Changes in the form of the oscillatory response, expressed by exponentially decaying sinusoidal components, can be used to identify sub-10 kHz changes in the RF frequency as well as RF amplitude and pulse arrival time. This makes Rydberg atom-based sensors promising for use as pulsed radar receivers, capable of identifying both the position and velocity of aircraft.

**ACKNOWLEDGEMENTS**

This project was supported by contributions from the Ontario Critical Technologies Initiative (CTI).

# REFERENCES

[1] C. S. Adams, J. D. Pritchard, and J. P. Shaffer, Rydberg atom quantum technologies, J. Phys. B At. Mol. Opt. Phys. **53**, 012002 (2019).
[2] J. A. Sedlacek, A. Schwettmann, H. Kübler, R. Löw, T. Pfau, and J. P. Shaffer, Microwave electrometry with Rydberg atoms in a vapour cell using bright atomic resonances, Nat. Phys. **8**, 819 (2012).
[3] J. A. Gordon, C. L. Holloway, A. Schwarzkopf, D. A. Anderson, S. Miller, N. Thaicharoen, and G. Raithel, Millimeter wave detection via Autler-Townes splitting in rubidium Rydberg atoms, Appl. Phys. Lett. **105**, 024104 (2014).
[4] H. Fan, S. Kumar, J. Sedlacek, H. Kübler, S. Karimkashi, and J. P. Shaffer, Atom based RF electric field sensing, J. Phys. B At. Mol. Opt. Phys. **48**, 202001 (2015).
[5] C. L. Holloway, M. T. Simons, J. A. Gordon, P. F. Wilson, C. M. Cooke, D. A. Anderson, and G. Raithel, Atom-Based RF Electric Field Metrology: From Self-Calibrated Measurements to Subwavelength and Near-Field Imaging, IEEE Trans. Electromagn. Compat. **59**, 717 (2017).
[6] S. M. Bohaichuk, F. Ripka, V. Venu, F. Christaller, C. Liu, M. Schmidt, H. Kübler, and J. P. Shaffer, Three-photon Rydberg-atom-based radio-frequency sensing scheme with narrow linewidth, Phys. Rev. Appl. **20**, L061004 (2023).
[7] M. Jing, Y. Hu, J. Ma, H. Zhang, L. Zhang, L. Xiao, and S. Jia, Atomic superheterodyne receiver based on microwave-dressed Rydberg spectroscopy, Nat. Phys. **16**, 911 (2020).
[8] D. A. Anderson, R. E. Sapiro, and G. Raithel, An Atomic Receiver for AM and FM Radio Communication, IEEE Trans. Antennas Propag. **69**, 2455 (2021).
[9] J. A. Sedlacek, A. Schwettmann, H. Kübler, and J. P. Shaffer, Atom-Based Vector Microwave Electrometry Using Rubidium Rydberg Atoms in a Vapor Cell, Phys. Rev. Lett. **111**, 063001 (2013).
[10] M. T. Simons, A. H. Haddab, J. A. Gordon, and C. L. Holloway, A Rydberg atom-based mixer: Measuring the phase of a radio frequency wave, Appl. Phys. Lett. **114**, 114101 (2019).
[11] C. L. Holloway, M. T. Simons, J. A. Gordon, and D. Novotny, Detecting and Receiving Phase-Modulated Signals With a Rydberg Atom-Based Receiver, IEEE Antennas Wirel. Propag. Lett. **18**, 1853 (2019).
[12] Y. Cai, S. Shi, Y. Zhou, Y. Li, J. Yu, W. Li, and L. Li, High-Sensitivity Rydberg-Atom-Based Phase-Modulation Receiver for Frequency-Division-Multiplexing Communication, Phys. Rev. Appl. **19**, 044079 (2023).
[13] S. Berweger, A. B. Artusio-Glimpse, A. P. Rotunno, N. Prajapati, J. D. Christesen, K. R. Moore, M. T. Simons, and C. L. Holloway, Closed-loop quantum interferometry for phase-resolved Rydberg-atom field sensing, Phys. Rev. Appl. **20**, 054009 (2023).
[14] S. J. Buckle, S. M. Barnett, P. L. Knight, M. A. Lauder, and D. T. Pegg, Atomic Interferometers, Opt. Acta Int. J. Opt. **33**, 1129 (1986).
[15] G. Morigi, S. Franke-Arnold, and G.-L. Oppo, Phase-dependent interaction in a four-level atomic configuration, Phys. Rev. A **66**, 053409 (2002).
[16] M. Schmidt, S. Bohaichuk, V. Venu, F. Christaller, C. Liu, F. Ripka, H. Kübler, and J. P. Shaffer, Rydberg-atom-based radio-frequency sensors: amplitude-regime sensing, Opt. Express **32**, 27768 (2024).
[17] H. C. Torrey, Transient Nutations in Nuclear Magnetic Resonance, Phys. Rev. **76**, 1059 (1949).
[18] Y. Li and M. Xiao, Transient properties of an electromagnetically induced transparency in three-level atoms, Opt. Lett. **20**, 1489 (1995).
[19] A. D. Greentree, T. B. Smith, S. R. de Echaniz, A. V. Durrant, J. P. Marangos, D. M. Segal, and J. A. Vaccaro, Resonant and off-resonant transients in electromagnetically induced transparency: Turn-on and turn-off dynamics, Phys. Rev. A **65**, 053802 (2002).
[20] D. A. Steck, *Quantum and Atomic Optics* (2007).
[21] P. R. Berman and R. Salomaa, Comparison between dressed-atom and bare-atom pictures in laser spectroscopy, Phys. Rev. A **25**, 2667 (1982).
[22] N. Lu, P. R. Berman, A. G. Yodh, Y. S. Bai, and T. W. Mossberg, Transient probe spectra in strongly driven atoms and their dependence on initial atomic conditions, Phys. Rev. A **33**, 3956 (1986).
[23] J. P. Shaffer and H. Kübler, A Read-out Enhancement for Microwave Electric Field Sensing with Rydberg Atoms, in Proc. SPIE Quantum Technologies, Vol. 10674 (2018).
[24] S. M. Bohaichuk, D. Booth, K. Nickerson, H. Tai, and J. P. Shaffer, Origins of Rydberg-Atom Electrometer Transient Response and Its Impact on Radio-Frequency Pulse Sensing, Phys. Rev. Appl. **18**, 034030 (2022).